%% file: paper4.tex
\documentstyle[]{l-aa}

\include{eps}

\begin{document}
\thesaurus{12.12.1, 11.04.1, 11.03.1, 11.09.5}
\title{Results of a search for emission-line galaxies towards nearby voids. The spatial
distribution.}
\author{Cristina C. Popescu \inst{1,3}
\and  Ulrich Hopp \inst{1,2}
\and  Hans Els\"asser \inst1
}

\offprints{Cristina C. Popescu}
\institute{Max Planck Institut f\"ur Astronomie, K\"onigstuhl 17, 
           D--69117 Heidelberg, Germany
\and Universit\"atssternwarte M\"unchen, Scheiner Str.1, 
       D--81679 M\"unchen, Germany   
\and The Astronomical Institute of the Romanian Academy, Str. Cu\c titul de
Argint 5, 75212, Bucharest, Romania}
\date{Received 20.03.1997; accepted 00.00.1997,}
\maketitle
\markboth{Cristina C. Popescu et al.: Results of a search for emission-line galaxies 
towards voids.}{}
\begin{abstract}

We present the results of a search for emission-line galaxies (ELGs) towards
nearby voids, based on a sample selected on the HQS (Hamburg Quasar
Survey)(Hagen et al. 1995)  - III-aJ objective-prism plates.  The survey was 
based on the presence of emission-lines and therefore has the advantage to 
detect very faint objects (with all the flux in the emission-lines and almost
 no continuum), that would be missed by an apparent magnitude limited 
survey. We found objects as faint as B=20.5 and M$_B=(-15.0,-12.0)$. The 
relevant 
brightness parameters of the sample are: (1) the sum of the flux in the 
emission-lines and of the continuum flux under the line, and (2) the 
equivalent widths (EW); the corresponding completeness limits are
 $6.7\times 10^{-14}\,{\rm erg}\,{\rm sec}^{-1}{\rm cm}^{-2}$ and 
8\,${\rm \AA}$, respectively. 

The observational data  are given in Popescu et al. (1996) and in 
the present paper we consider a complete subsample from these data. We 
analyse the 
spatial distribution of our sample of ELGs in comparison with the distribution
of normal galaxies. Overall both  distributions trace the same structures. 
Nevertheless we also found a few ELGs in voids, from which at
least 8 lie in the very well defined nearby voids. The isolated galaxies seem
to form fainter structures that devide the larger voids into smaller ones.
 From our estimates of the
expected number of void galaxies we conclude that  we
did not find an underlying homogenous void population.
 
\end{abstract}
\section{Introduction}

The large scale structure of the Universe, as derived from recent redshift
surveys, reveals us large inhomogeneities: we
see the galaxies forming clusters and superclusters, or even 
larger structures,  filaments and sheets of galaxies, or walls, which 
surround the empty regions that are the voids. But this view is
primary based on the high luminosity galaxies - the giants. There has been a
long debate in the literature whether the giant galaxies are fair tracers of
the large scale structure and whether the voids are really empty. From an
observer point of view, the galaxy maps may reflect special observational
selection effects in surface brightness, integral magnitude or diameter. It is 
a straightforward idea to think that going fainter we will eventually start to 
fill the voids. We know that the dwarf
galaxies dominate the galaxy number in the Universe, but only very few are
contained in the present galaxy catalogues. These objects can be very small,
intrinsically faint and can also have very low surface brightness. The dwarf 
galaxies are thus good candidates to fill up the voids. There are also 
some theories of galaxy formation (Dekel \& Silk 1986)
that were worked out in the frame of the cold dark matter (CDM) scenarios 
including biasing, which predict that the dwarf galaxies should originate from 
the 1\,${\sigma}$ fluctuations, expecting thus to be more evenly distributed 
than the high rare density peaks which form the giants. In these scenarios the 
dwarf galaxies should trace the underlying dark matter and are expected to 
fill the voids. It is therefore a fundamental task to search for the existence 
of a homogenous void population. Whether or not such a population exists  
would be of extreme importance for our understanding of galaxy formation and 
large scale structure formation and evolution.

Several studies of the spatial distribution of dwarf galaxies  were carried out
to answer these questions. Different projects were conceived to 
search for galaxies in voids or to see if the clustering properties of dwarfs 
galaxies are different from that of the normal giant galaxies.  Most of the 
studies were based on redshift surveys of low-surface brightness dwarfs 
(Bothun et al. 1986,
Thuan, Gott \& Schneider 1987, Eder et al. 1989, Salzer, Hanson \& Gavazzi 
1990, Binggeli, Tarenghi \& Sandage 1990, Weinberg et al. 1991,  Thuan et
al. 1991). The results of these studies
did in general not favour biased galaxy formation and no void population
was found. More recently, a new search for faint galaxies in voids was 
accomplished, with the intention to
overcome the limitations of previous surveys in magnitude, diameter and surface
brightness (Hopp 1994, Hopp et al. 1995, Kuhn et al. 1997). Other studies were
based on emission-line objects (Tifft et al. 1986, Moody et al. 1987, Weistrop
\& Downs 1988, Weistrop 1989, Salzer 1989, Rosenberg et al. 1994) and there
has been the suggestion that a few emission-line galaxies have been found in
the voids.  Much effort has concentrated on the study of the Bootes void, a
huge low density region with a volume of $1.3\times 10^{5}\,{\rm Mpc}^{3}$,
first discovered by Kirshner et al. (1981) (see also Kirshner et al. 
1983\,a,b). Up to now 58 galaxies were
identified in the Bootes void, most of them coming from objective-prism
surveys (Sanduleak \& Pesch 1982, Tifft et al. 1986) but also from 
IRAS (Dey et al. 1990) and HI surveys (Szomoru et al. 1996a). The studies of 
the properties of these galaxies (Weistrop et al. 1995, Szomoru et al. 1996b) 
showed that they were strong emission-line objects with significant amount of 
star formation. They were also luminous galaxies, not the low mass, low surface
brightness galaxies predicted to be found in voids. On the other hand it is
important to notice that the Bootes void is  beyond the distance at which
the large scale structure is well defined by the present catalogues.

Having in mind that no definitive conclusion was drawn from previous
studies and no faint void population was found in literature, we have
undertaken a programm to search for dwarf galaxies in
voids (Popescu et al. 1995, Popescu 1996). We especially selected nearby voids which were very well defined
in the distribution of normal galaxies, in order to overcome some of the
limitations of previous surveys like those in Bootes. Since there was a 
hint that emission-line galaxies were found in
voids, we chose to search for emission-line objects, but with the aim of 
finding mainly dwarf HII galaxies or BCDs. The observational 
data of our survey as well as the location of our void regions are given in
 Popescu et al. (1996) (Paper 1).  The candidates were selected on the
objective prism plates taken from the Hamburg Quasar Survey (HQS)(Hagen et
al. 1995) and therefore they have the name built with the prefix HS. 

 In this paper we will give the results of our search for galaxies in voids 
and we will describe the spatial distribution of the ELGs 
in comparison with the normal galaxies. The paper is organised as follows. In
section 2 we estimate the completeness of our survey in terms of fluxes
and equivalent widths. In section 3 we give a qualitative description of the 
spatial distribution of our ELGs, by using cone-diagrams in both 
Right-Ascension and Declination
projection, and we also quantify the results by calculating the spatial
densities and the nearest neighbour distributions. In 
section 4 we analyse the main properties of the galaxies we found in voids
and in the last section we summarise and give the conclusions.

\section{The completeness of the survey}

In Paper 1 we have discussed in detail the selection 
 of our sample and the selection effects. In summary, an automated
procedure was applied to the low-resolution digitised objective-prism spectra, 
based on two parameters, the slope of the continuum and the 
\lq\lq luminosity\rq\rq  
of the integrated spectra (in counts). The selected candidates were 
afterwards
rescanned with high resolution and the final selected spectra were visually
inspected for the presence of emission-lines. While the slope of the continuum 
helps us to preselect very promising emission-line candidates, the cut in  
brightness at the faint end of
the photographic plates produces some loss of very faint objects, with very
little continuum and almost all the flux in the emission-lines. In order to
prevent the latter incompleteness we also scanned the faint end of the
photographic plates, and we completed the follow-up spectroscopy for all the 
faint candidates. This 
extra survey was done only for one of our regions - Region 3 from Paper 1 
(a region 
North to the Coma Supercluster, $30.5^{\circ}<{\delta}<45.5^{\circ}$, centred 
around $13.5^{\rm h}$)
and in the following discussion we will refer only to this subsample. The
surface density of the subsample is 0.3 galaxies/deg$^{2}$ and the
catalogue of the additional survey will be published elsewhere. For the data
analysed in this paper  we have 
completed the spectrophotometry, allowing thus to establish
a completeness limit. Our sample was not selected in a traditional way,
therefore it is not a continuum magnitude limited sample. Nevertheless we can 
first give the limits of our survey in continuum blue magnitudes, based on 
the data from Paper 1 and from the additional survey. Our sample contains
 objects as faint as B$_{lim}=20.5$, and also intrinsically faint 
objects, down to M$_{B}=(-15.0,-12.0)$. This indicates that the present survey goes
deeper than other surveys and is therefore adequate for a search for faint
objects in voids. But for a sample that was selected based on the
presence of the emission-lines, the relevant brightness parameters are the sum
of the flux in the emission-lines and of the flux in the continuum under the
line, and the equivalent widths. A detailed description of this selection
procedure is given by Salzer (1989). 

As the main selection was based on the
presence of the [OIII] ${\lambda}$5007 line, the corresponding parameters for 
this line were computed. A complete
catalogue with fluxes and equivalent widths (EW) will be published in a 
following paper. All the 
spectroscopic parameters calculated here are
$4^{\prime\prime}$ slit widths measurements. The line flux F$_{L}$ was measured
directly from the slit spectra. The flux in the continuum under the line was
calculated as ${\rm F}_C  =  {\rm f}_c\,{\Delta}{\lambda}$, 
where f$_{c}$ is the mean flux per unit of wavelength and ${\Delta}{\lambda}$
 is the FW0I (flux width at zero intensity) of the emission-line. 
f$_{c}$ is calculated as the ratio between 
the line flux and the EW of the line, ${\rm f}_{c}={\rm F}_{L}/{\rm EW}$ and 
it can be measured directly from the slit spectra.

${\Delta}{\lambda} =2\,{\rm Disp}({\rm z}){\rm  R}$, where Disp(z) is the reciprocal 
dispersion of the
objective prism in ${\rm \AA}\,{\rm mm}^{-1}$ and R is the spectral resolution on 
the objective
prism plates. In our case the resolution R is determined by the slit widths of
the PDS machine that was used to digitise the plates. For the high resolution
scans (see Paper 1 for further details) we used a slit of 
0.03\,mm and we
can assume that this is also the value of R\protect\footnote{Because we selected only
high quality plates, the broadening of the lines
by seeing is almost negligible compared to the PDS slit width, as 0.03\,mm
corresponds to about 2$^{{\prime}{\prime}}$.}. The resulting 
${\rm F}_{L+C} = {\rm F}_{L}+{\rm F}_{C}$
as well as the EW of the [OIII] ${\lambda}$5007 lines were computed for each 
individual object of our sample. For two cases the H$_{\beta}$ was stronger 
than the [OIII] lines,
indicating starburst like galaxies, and
therefore H${\beta}$ was measured instead.

\begin{table}[htp]
\caption[]{The V/V$_{m}$ test. }
{\footnotesize
\begin{tabular}{rcrrr}
                 & & & &  \\
\hline\hline
                 & & & & \\
 m$_{L+C}$ & ${\rm V/V}_{m}$ & N & N$^{\prime}$ & c$\%$ \\
(1) & (2) & (3) & (4) & (5) \\ 
                      & & & & \\
\hline
                      & & & &\\
 10.8 & 0.581 & 12 &  0 & 100.0 \\
 10.9 & 0.597 & 16 &  0 & 100.0 \\
 11.0 & 0.575 & 19 &  0 & 100.0 \\
 11.1 & 0.536 & 21 &  0 & 100.0 \\
 11.2 & 0.502 & 23 &  0 & 100.0 \\
 11.3 & 0.527 & 29 &  0 & 100.0 \\
 11.4 & 0.473 & 30 &  2 &  93.3 \\
 11.5 & 0.477 & 35 &  1 &  92.1 \\
 11.6 & 0.452 & 38 &  3 &  86.4 \\
 11.7 & 0.459 & 44 &  2 &  84.6 \\
 11.8 & 0.464 & 51 &  1 &  85.0 \\
 11.9 & 0.453 & 57 &  4 &  81.4 \\
 12.0 & 0.433 & 62 &  6 &  76.5 \\
 12.1 & 0.414 & 67 &  8 &  71.2 \\
 12.2 & 0.389 & 71 & 11 &  65.1 \\
 12.3 & 0.368 & 75 & 13 &  59.5 \\
 12.4 & 0.334 & 77 & 17 &  53.1 \\
 12.5 & 0.327 & 82 & 17 &  49.1 \\
 12.6 & 0.324 & 88 & 20 &  45.6 \\
 12.7 & 0.283 & 88 & 29 &  39.6 \\
 12.8 & 0.273 & 92 & 30 &  35.9 \\
 12.9 & 0.258 & 95 & 35 &  32.3 \\
 13.0 & 0.225 & 95 & 44 &  28.1 \\
 13.1 & 0.203 & 96 & 50 &  24.7 \\
 13.2 & 0.184 & 97 & 57 &  21.7 \\
 13.3 & 0.167 & 98 & 66 &  19.1 \\
 13.4 & 0.146 & 98 & 76 &  16.6 \\
 13.5 & 0.134 & 99 & 87 &  14.6 \\

      &       &    &    &            \\
\hline
\end{tabular}
}
\end{table}

The F$_{L+C}$ was transformed in a magnitude scale:
\begin{eqnarray}
 {\rm m}_{L+C} & = &
 -2.5 \times log10({\rm F}_{L+C}) - 20.9,
\end{eqnarray}
 where the constant is arbitrary. 

Our \hspace{-0.05cm}sample contains objects with the flux of the 
[OIII] ${\lambda}$5007 emission-line as faint as 
$5.3 \times 10^{-15}\,{\rm erg}\,{\rm sec}^{-1}\,{\rm cm}^{-2}$ and as bright 
as $5.3 \times 10^{-13}\,{\rm erg}\,{\rm sec}^{-1}\,{\rm cm}^{-2}$. If we 
consider the brightness parameter discussed above, namely the sum of the 
flux in the emission-line and
of the continuum under the line, then the range is
$9.5 \times 10^{-15}{\rm erg}\,{\rm sec}^{-1}\,{\rm cm}^{-2}<{\rm F}_{L+C}<
8.0 \times 10^{-13}{\rm erg}\,{\rm sec}^{-1}\,{\rm cm}^{-2}$. The EW values 
range between 8\,${\rm \AA}$ and 1700\,${\rm \AA}$. 

The completeness limit was derived based on a V/V$_{m}$ test (Schmidt 1968). 
V is the volume contained in a sphere whose radius is the (redshift) distance
to the object and V$_{m}$ is the volume contained in a sphere whose radius is the
maximum distance the galaxy could have and still be in the sample under study,

\begin{eqnarray}
{\rm V}_{m} & = & \frac{4}{3}{\pi}\times 10^{0.6({\rm m}_{lim}-{\rm M}-
{\rm A}-25.0)}
\end{eqnarray}

where m$_{lim}$ is the completeness limit and A is the Galactic absorption.
 
The value of V/V$_{m}$ is then given by $10^{0.6(m-m_{lim})}$. The mean 
value of the ratio V/V$_{m}$
should be 0.5 for a complete sample of objects uniformly distributed in
Euclidian space. In practice the distribution of galaxies is affected by large
scale structure inhomogeneities. As a first approximation we can consider that
our subsample covers enough volume (415 deg$^2$, 
v$\le 15000\,{\rm km}/{\rm s}$) to cancel out these 
effects. Also we
will show that the ELGs have a small tendency to be more evenly distributed
than the giant galaxies, lying also in some voids or at the rim of the voids,
and 
therefore the approximation of uniformity can be applied as a first guess. 

The mean V/V$_{m}$ ratios 
were computed for 99 galaxies in our
Region 3 and the results are listed in Table 1. The Column (1) gives the 
m$_{L+C}$,
Column (2) gives the V/V$_{m}$ ratios and Column (3) gives the total number of 
objects brighter than the corresponding m$_{L+C}$. Column (4) specifies the
number of objects that need to be added at each level of magnitude in order to
keep the average V/V$_m$ around 0.5 and Column (5) gives the level of
completeness, c$\%$.
The V/V$_{m}$ ratios are around 0.5 up to m$_{L+C}=11.4$ and then they start
 to decrease. We will take as a  completeness 
limit m$_{L+C}=12.0$, where the
sample is 77$\%$ complete. This corresponds to a flux of 
$6.7\times10^{-14}$erg\,sec$^{-1}{\rm cm}^{-2}$. 

\begin{figure}[htb]
\plotfiddle{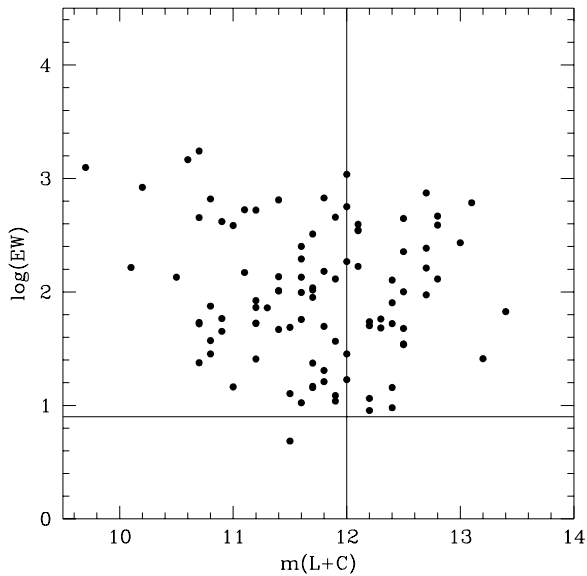}{4.0in}{0.}{40.}{40.}{-120}{-30}
{\baselineskip 1.5cm \caption[]{ A plot of log(EW)
versus m$_{L+C}$, where EW are the equivalent widths in ${\rm \AA}$ and
m$_{L+C}$ is the flux in the emission-line plus the flux in the continuum under
the line, transformed in a magnitude scale (see (1)).  With the vertical line we delimit the 
complete sample from 
the incomplete one and with the horizontal line we trace the threshold below 
which the ELGs are no more seen by our survey.   }}
\end{figure}

In order to determine also the EW limit of our survey we plotted in Figure 1
 log(EW)
versus m$_{L+C}$. With the vertical line we delimit the complete sample from 
the incomplete one and with the horizontal line we trace the threshold below 
which the ELGs are no more seen by our survey. This is a level of 0.9, which 
means an ${\rm EW}=8{\rm \AA}$. There is only one point that falls below the
horizontal threshold of 0.9. The corresponding 
galaxy was selected based on its [OII] $\lambda$3727 line, one of the few cases
 that did not use only the [OIII] ${\lambda}$5007 line criteria. Its spectrum 
is typical for a low 
ionization object, with faint [OIII] ${\lambda}$5007 and strong [OII] 
${\lambda}$3727 emission lines. 
If one computed the EW for the [OII] ${\lambda}$3727 line, the galaxy would 
fall above the horizontal threshold. One should also mention that all the
points that were just above this threshold were galaxies selected as
second priority objects (see Paper 1 for a detailed discussion of the
selection procedure). The corresponding emission-lines were barely 
detectable on the digitised spectra, and we had difficulties to decide 
whether the candidate was real or
not. The follow-up spectroscopy was the only method to determine the real 
nature of the objects. Therefore, removing these points from the plot, the 
diagram would indicate a slightly higher limit in EW, toward 12\,${\rm \AA}$.

The diagram also shows a trend of increasing EW at both the bright and the 
faint end of the m$_{L+C}$. At the very bright end the galaxies have a strong
continuum, and therefore it requires a higher EW for the emission-line to be
detected above the continuum. By contrary, at the faint end, the galaxies 
have a low level continuum and therefore the spectrum is quite noisy. It 
requires again a high EW for the emission-line to be detected above the noise. 
In addition the sum between the flux in the emission-line and the continuum 
flux has to be kept above a certain level of detectability, and as the
 continuum decreases, the flux in the emission-line has to increase in order 
to detect the galaxy.   

In conclusion we can build a complete sample with all the objects brighter 
than F$_{L+C}\geq 6.7\times 10^{-14}$erg\,sec$^{-1}{\rm cm}^{-2}$ and having a
detectability in equivalent widths 
EW$\geq 8{\rm \AA}$. Such a sample cannot be compared with a magnitude 
selected sample, but can be used for
statistical calculations. 
 
\section{The distribution of ELGs}
\subsection{The space densities}

We calculate the space densities of our sample of ELGs using a 
V/V$_m$ method applied to the parameter m$_{L+C}$, as defined in the previous
section.  Through this paper we used a Hubble constant 
${\rm H_0}=75$\,km/s/Mpc.  We computed a corresponding \lq\lq absolute 
magnitude\rq\rq, which is 
the intrinsic flux of the emission-lines plus the continuum under the line,
transformed in a magnitude scale, 
M$_{L+C}$. We should once again mention that these so called absolute
magnitudes do not have the meaning of the continuum absolute magnitudes and 
should be interpreted as the intrinsic strength of the emission-lines. The 
space density at M$_{L+C}$ is :

\begin{eqnarray}
{\Phi}({\rm M}_{L+C})=\frac{4\,{\pi}}{{\Omega}}\sum_{i}(\frac{1}{{\rm V}_m^i})
\,\,\,\,\,{\rm Mpc}\,^{-3}\,{\rm mag}^{-1}
\end{eqnarray}

where ${\Omega}$ is the solid angle covered by our survey, and the summation is
over all galaxies in the absolute magnitude interval M$_{L+C}\pm$0.5\,mag. The 
absolute magnitudes were calculated considering a Galactic absorption given by
 ${\rm A} = 0.23/(sin|{\rm b}|)$.

\begin{table}[htp]
\caption[]{Space Densities $\Phi({\rm M}_{L+C})$. }
\begin{tabular}{rcr}
                 & &  \\
\hline\hline
                 & &  \\
 M$_{L+C}$ & $log{\Phi}({\rm M}_{L+C})$ & N \\
(1) & (2) & (3)\\ 
                      & &\\
\hline
                      & &\\

-27.0 & -6.34 &  2 \\ 
-26.0 & -5.72 &  3 \\ 
-25.0 & -4.40 & 17 \\ 
-24.0 & -3.78 & 18 \\ 
-23.0 & -3.13 & 21 \\
-22.0 & -2.86 &  9 \\
-21.0 & -2.48 &  6 \\
-20.0 & -2.27 &  3 \\
-19.0 & -1.46 &  3 \\
  
& &\\
\hline

\end{tabular}
\end{table}

In the computation of the $\Phi({\rm M}_{L+C})$ 
we included
all galaxies up to a m$_{L+C}=12.5$, for which the completeness level is
49.1$\%$. Thus we must increase the calculated space densities by a factor of 
2.04 to allow for incompleteness. In Table 2 we listed the
 $log{\Phi}({\rm M}_{L+C})$ for each bin of absolute magnitude together with the
number of galaxies included in each bin.

Table 2 shows that we found only a few extreme strong 
[OIII] ${\lambda}$5007 line objects (5$\%$) and that most of our objects
(57$\%$) have high and intermediate strengths of the emission-lines. Going to 
objects that have intrinsically faint [OIII] ${\lambda}$5007 lines, our survey 
becomes less efficient, and some of these objects can be better detected by 
H${\alpha}$ surveys (see Zamorano et al. 1994).
 
The integration over the whole range of absolute magnitudes M$_{L+C}$ gives a 
space density of 0.046$\pm0.005\,{\rm Mpc}^{-3}$. In order to calculate the 
corresponding error we considered different completeness limits and we 
estimated the spread around the assigned value. The integrated space density 
is 
nevertheless dominated by the last bin (at the faint end) of the luminosity 
function, which
was calculated based only on three galaxies. This point is therefore very
uncertain and we prefer to give a space density integrated only till
M$_{L+C}=-20.0$. Then we obtain $\Phi=0.011\pm0.001\,{\rm Mpc}^{-3}$, which is
a factor 3.5 lower then the previous value. The results for different 
completeness limits are now more stable, with the estimated error a factor 5 
lower.

\subsection{Cone-diagrams}

A study of the spatial distribution of the ELGs requires a comparison with a
catalogue of normal galaxies that would properly trace the main structures in
the nearby Universe and would also properly define the nearby voids.  We have
already mentioned in the introduction that we selected our surveyed regions to
contain well defined nearby voids in the distribution of the giant galaxies.
For the comparison catalogue we used the latest electronic version of the 
 ZCAT (April 1995)\protect \footnote{ For the description of the catalogue
 see also Huchra et al. 1992, Huchra et al. 1995.} and we selected all galaxies
 brighter than B=15.5. The ZCAT contains only one strip that is complete 
 to $15.5^{m}$. This is the so called
\lq \lq Slice of the Universe\rq\rq, $26.5^{\circ}< {\delta} < 32.5^{\circ}$
 and $8^{\rm h}<{\alpha}<17^{\rm h}$.  Our surveyed regions are outside the 
zone of the Slice,
therefore the catalogue is not complete to $15.5^{m}$. The incompleteness of
the comparison catalogue do not affect the qualitative description of the
large-scale structure, as given by the cone-diagrams, but could affect the
results of some statistical tests, as mentioned in the next section.

The most common way to visualise the spatial distribution of galaxies is to 
use cone diagrams in both redshift-Right Ascension and in redshift-Declination 
plane. Since we give a qualitative description of the 
spatial distribution, we have plotted all ELGs, without respect to their
completeness. Such a restriction is done only when we 
apply statistical tests (see subsection 3.3). The diagrams contain also the
comparison catalogue (the ZCAT galaxies brighter than 15.5). In addition we have
plotted the ZCAT galaxies fainter than 15.5, with the intention to have a 
first impression of how the 
fainter galaxies 
start to structure and how they correlate with the ELGs. In the following 
description we will use the
term bright ZCAT for the ZCAT galaxies with apparent magnitudes brighter than
B=15.5 and the term faint ZCAT for the ZCAT galaxies fainter than B=15.5. The 
terms
bright and faint do not therefore refer to the intrinsic brightness of the
comparison galaxies. As most 
of our galaxies are fainter than 15.5, there is practically no overlap in 
apparent magnitudes between our sample and the comparison (bright) catalogue. 

All velocities are corrected 
for the Galaxy's 
motion with respect to the velocity centroid of the Local Group. We use a 
correction term of ${\Delta}{\rm v}=300\,sin({\rm l})\,cos({\rm b})$ 
(Sandage 1975), 
where ${\rm l}$ and ${\rm b}$ are the galactic coordinates. We plotted all the 
galaxies with  velocities out to $15000\,$km/s, which corresponds to a redshift
of ${\rm z}=0.05$, including thus most of our sample.  For velocities greater 
than 15000\,km/s the comparison catalogue becomes essentially non-existent.
 Already for velocities greater than 10000\,km/s the ZCAT quickly thins out,
 and therefore we refrain from drawing conclusions about the reality of voids 
beyond 10000\,km/s.

In the cases where our surveyed region was too narrow in one dimension or not
properly covering some main features of the large scale structures, a
bigger area of sky is displayed. In these cones the surveyed region is 
either delimited with dotted lines or some specifications are given in the
Captions. We should also mention that our cone diagrams do 
not completely cover all the surveyed region, and the strips are chosen to
 display the most relevant information. 
\begin{figure*}[hp]
\plotfiddle{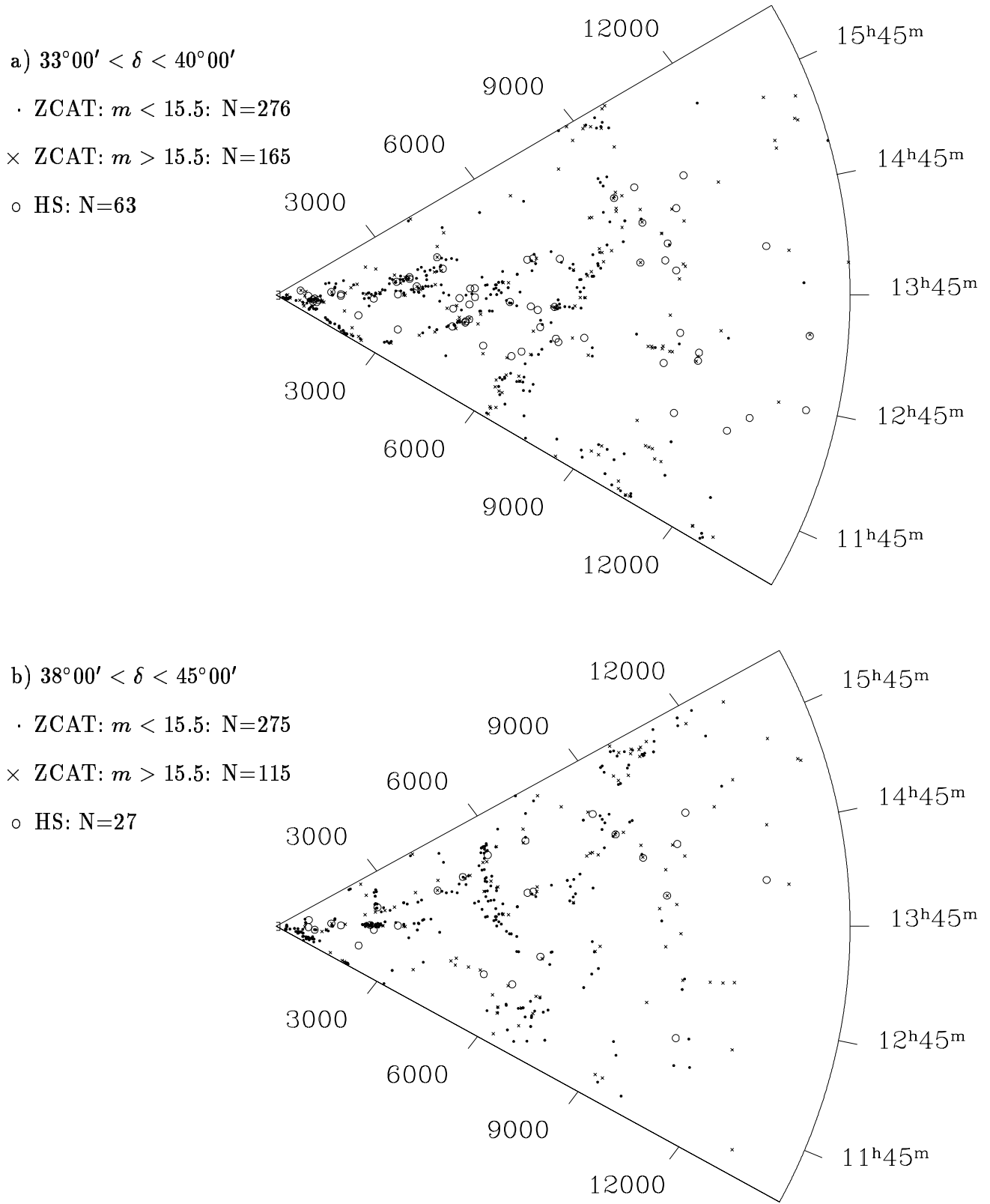}{8.5in}{0.}{100.}{100.}{-280}{-190}
\vspace{0.8cm}
{\baselineskip 1.5cm \caption[]{Wedge-plots of redshift (cz in km/s) - right 
Ascension out to a redshift of 15000\,km/s. The ZCAT $m<15.5$: small dots, the
ZCAT $m>15.5$: crosses, the ELGs: open circles. 
The wedge is a $7^{\circ}$ wide
strip in Declination centred on a) ${\delta}=36^{\circ}30^{\prime}$, 
b) ${\delta}=41^{\circ}30^{\prime}$. The cones
contain a larger angle in right Ascension than our surveyed region, which is
only between ($12^{\rm h}$, $15^{\rm h}30^{\rm m}$). }}
\end{figure*}
 The cones plotted in Fig. 2\,a,b contain
a larger angle in Right Ascension than our surveyed region, because we did not
 want to cut through the very well defined foreground voids and we wanted
to have a better impression of the large scale structure in this region. The
actual surveyed region is only between 12$^{\rm h}$ and 
15$^{\rm h}$30$^{\rm m}$.

The strip plotted in Fig. 2\,a is slightly to the North of the 
\lq\lq Slice of the Universe\rq\rq.
For this reason we can still see some of the prominent features of the  
Slice,
remnants of the \lq\lq Harvard Sticky Man\rq\rq\ at velocities less than 
7500\,km/s, but
without the Coma Cluster. The most remarkable feature of the diagram is 
the
\lq\lq Great Wall\rq\rq, which crosses our diagram from 
$6500\,$km/s at
${\alpha}=12^{\rm h}00^{\rm m}$ to $9000\,$km/s at 
${\alpha}=15^{\rm h}30^{\rm m}$. The structures of 
the Sticky Man
define some foreground voids, of which that centred at
${\alpha}\sim13^{\rm h}15^{\rm m}$, ${\rm v}\sim3000\,$km/s is one of the best
defined void in our surveyed region and we will call it Void 1. In the front 
of the
Great Wall there is a very big void that opens to the western side of the cone
and continues also outside our actual survey. We will call it Void 2. Void 1
and Void 2 will be used to draw our conclusions about the void population. 
Beyond the Great Wall there is
 another  big void, but 
the size of the void is no more well defined by the ZCAT galaxies because 
at these distances  the comparison catalogue becomes practically non-existent.

\begin{figure*}[hp]
\plotfiddle{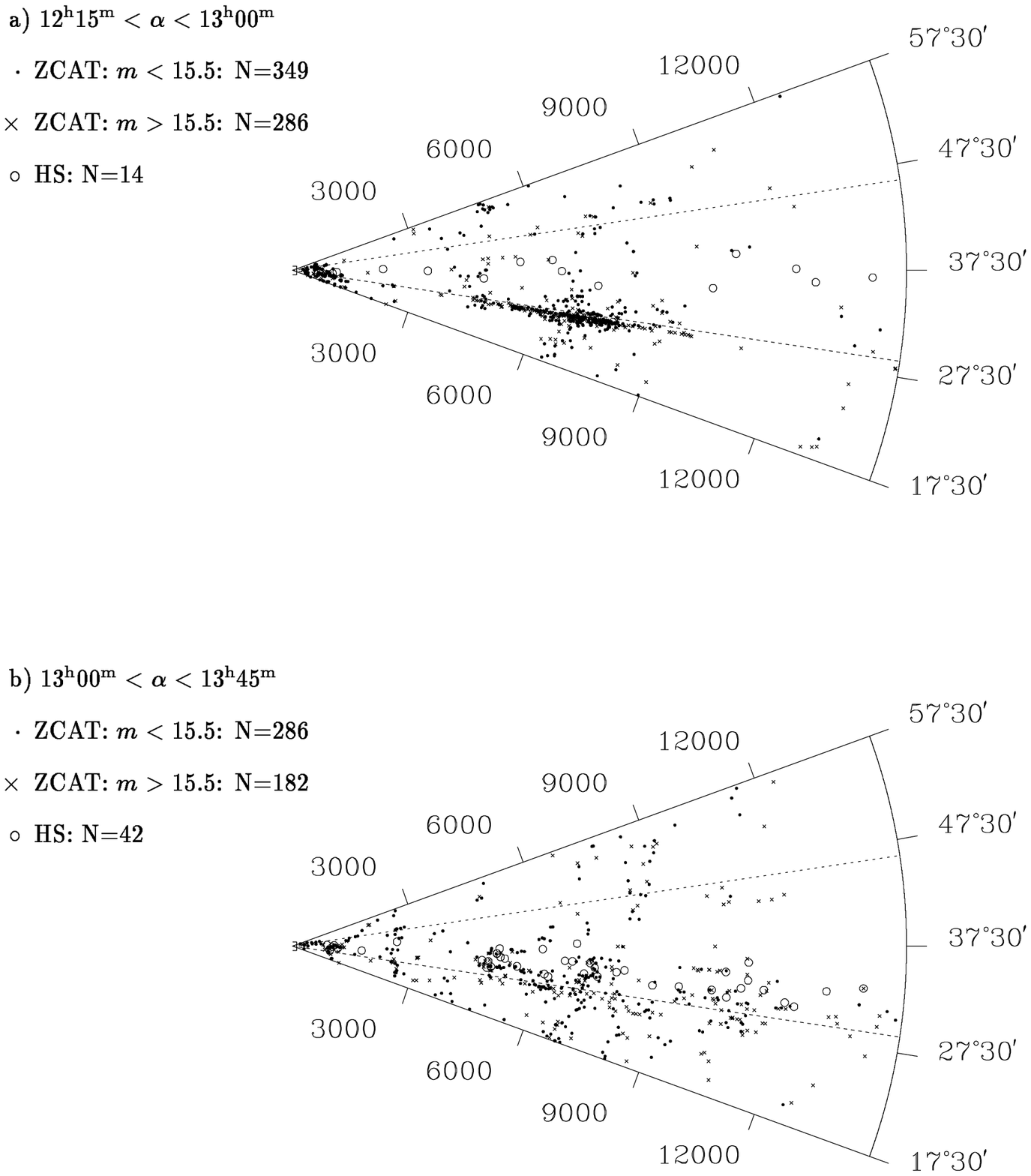}{8.5in}{0.}{100.}{100.}{-280}{-190}
{\baselineskip 1.5cm \caption[]{Wedge-plots of redshift (cz in km/s) -
Declination  out to a redshift of 15000\,km/s. 
The ZCAT $m<15.5$: small dots, the
ZCAT $m>15.5$: crosses, the ELGs: open circles. The wedge is a $45^{\rm m}$ wide
strip in Right Ascension. }}
\end{figure*}
\addtocounter{figure}{-1}
\begin{figure*}[hp]
\plotfiddle{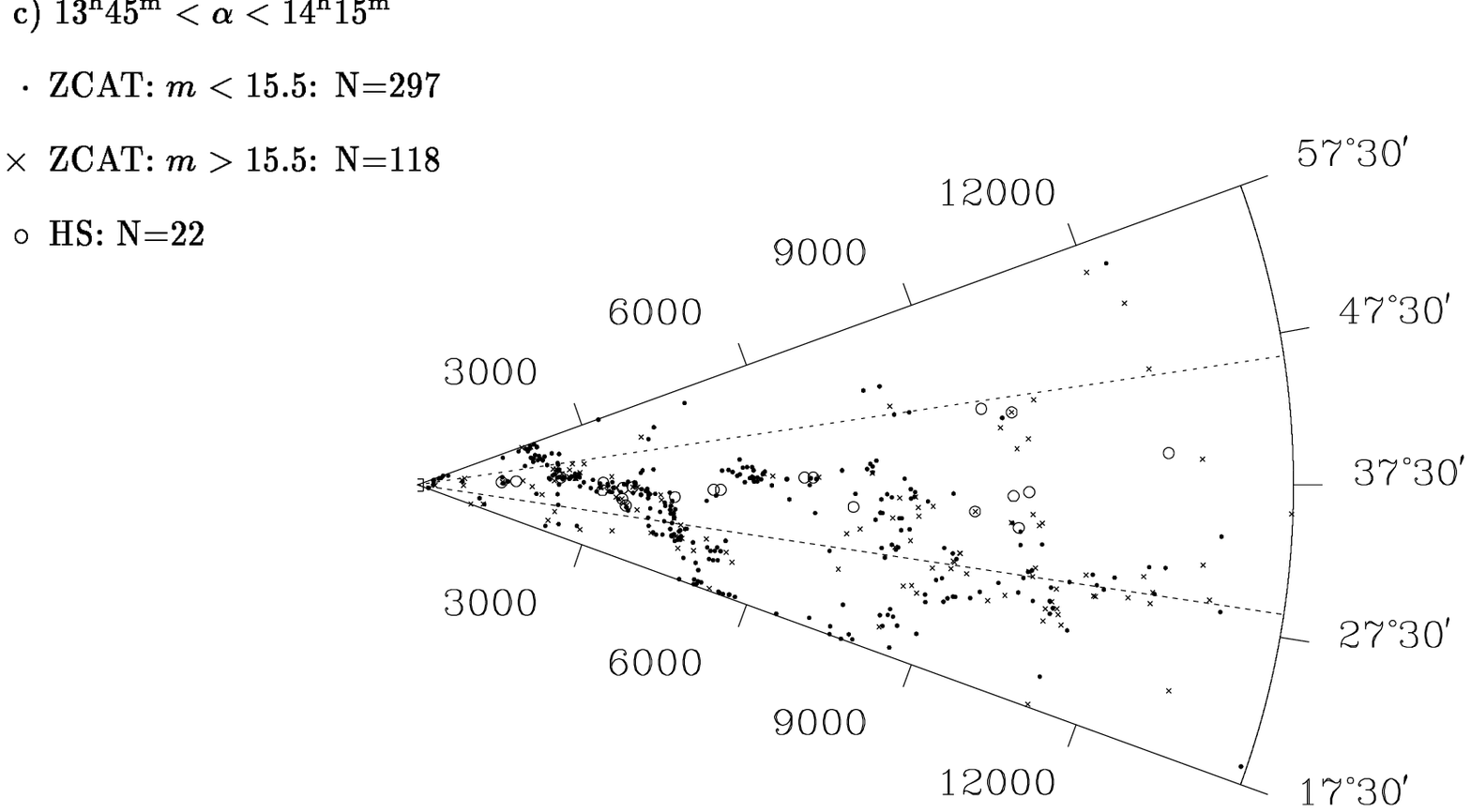}{8.5in}{0.}{100.}{100.}{-280}{-190}
{\baselineskip 1.5cm \caption[]{Continued  }}
\end{figure*}
\begin{figure*}[hp]
\plotfiddle{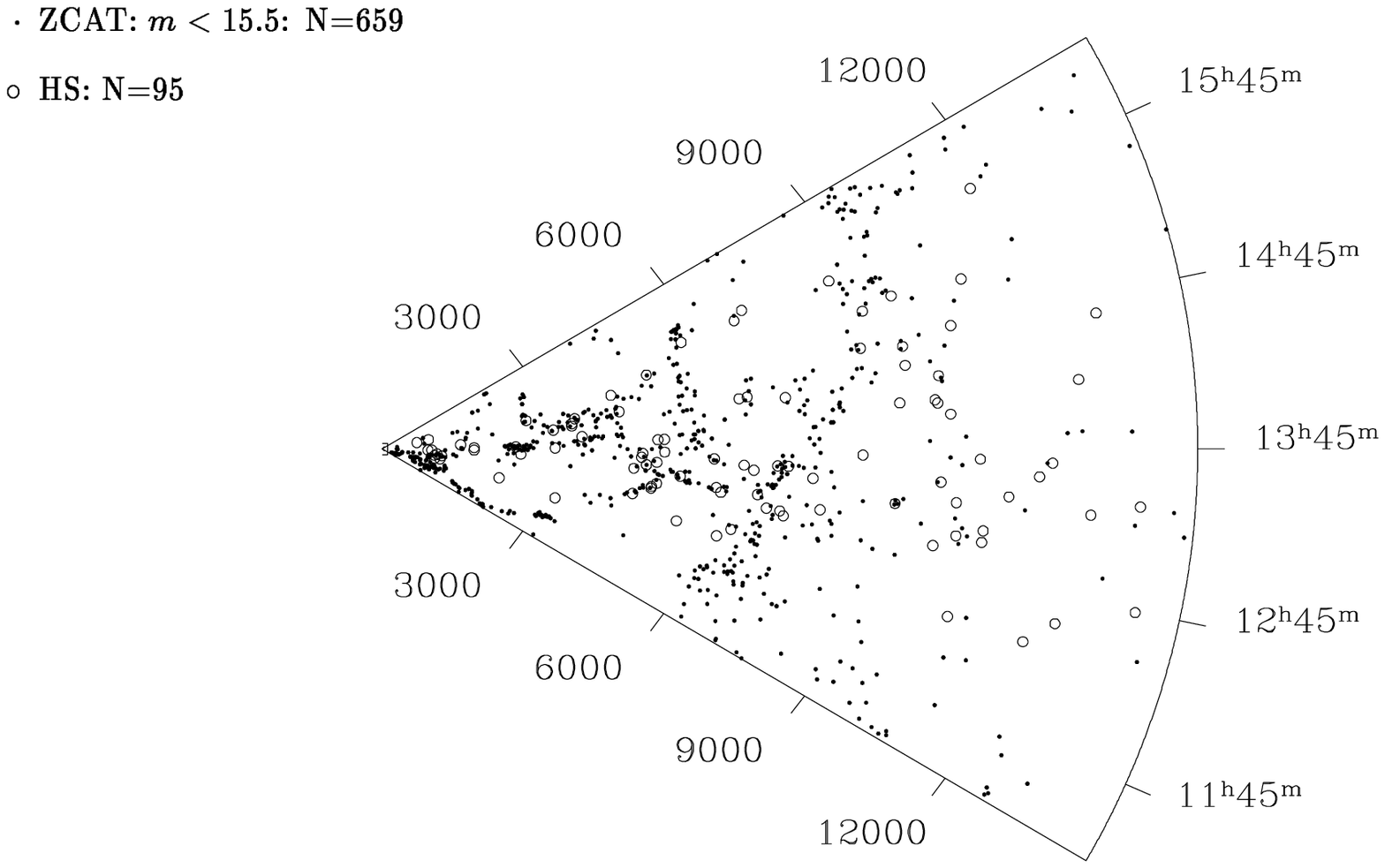}{8.5in}{0.}{100.}{100.}{-280}{-190}
{\baselineskip 1.5cm \caption[]{Wedge-plots of redshift (cz in km/s) -
 Right-Ascension  out to a redshift 
of 15000\,km/s. The ZCAT $m<15.5$: small dots, the
ZCAT $m>15.5$: crosses, the ELGs: open circles. 
The wedge is a $15^{\circ}$ wide strip in Declination.  }}
\end{figure*}

Our galaxies seem to follow the structures described above as well. At a closer
inspection one can discover that there are some galaxies that lie very isolated
in some of the foreground voids. In Void 1
there are two galaxies, HS1236+3821, ${\rm v}=2215$\,km/s and HS1226+3719,
${\rm v}=3306$\,km/s, that have the nearest bright ZCAT galaxy at a distance of
 $3.85\,h^{-1}$\,Mpc and $6.63\,h^{-1}$\,Mpc, respectively (for detailed 
description of 
the isolated galaxies see Table 3). At this distance the mean separations
between galaxies is around $0.5\,h^{-1}$\,Mpc, so these two galaxies are 
extremely isolated. They are among the best candidates we found in the voids.
Two further faint ZCAT galaxies are also present in the void (see Table 4). 
In Void 2  
we found an \lq\lq Arch\rq\rq of 7 ELGs (HS1236+3937, HS1232+3947, 
HS1240+3721, HS1332+3426, HS1328+3424, HS1310+3801), that seem to divide the 
void into three smaller voids. The galaxy HS1236+3937 has the largest 
isolation, of
$8.68\,h^{-1}$\,Mpc. The Arch is also populated  by three faint ZCAT galaxies
 while a
fourth one closes the Arch at lower redshifts (Table 4).
In the 
background void beyond the Great Wall there are
two HS galaxies, one at 8131\,km/s, HS1306+3320, and one at ${\rm v}=9558\,$km/s,
HS1410+3446. However we have already
mentioned that this background void is not delimited at the far distance edge
 by the bright ZCAT galaxies, but mainly by our
ELGs and by some faint ZCAT galaxies. It is anyway remarkable the large number of ELGs 
we found at higher redshifts, where the ZCAT catalogue do not contains any 
galaxy.

Fig. 2\,b displays the same region in Right Ascension but shifted to the North 
and
having an overlap with the previous diagram of 2$^{\circ}$ in Declination. We
chose a small overlap in order to see how the structures evolve when moving in
one coordinate. The diagram no longer resemble the Slice, though 
one can
still see some features of the Great Wall. There is an extra feature that
appears at $6000\,$km/s, a filament that stretches from 
${\alpha}=13^{\rm h}45^{\rm m}$
 up to ${\alpha}=15^{\rm h}40^{\rm m}$. The two voids in the front
of the Great Wall seem to converge now in a unique void centred on
$\sim4000\,$km/s. The void still contains some galaxies from the Arch, but one
 can see now a chain of faint ZCAT galaxies that seem to associate with the ELGs
and again divide the big void in two smaller voids. The region
beyond ${\rm v}=12500\,$km/s and east to $13^{\rm h}30^{\rm m}$ contains the 
southwest boundary 
of the Bootes void (centred on ${\alpha}=14^{\rm h}50^{\rm m}$, 
${\delta}= 46^{\circ}$, ${\rm v}=15500\,$km/s) (Kirshner et al. 1981).

In Fig. 3\,a we plotted in a redshift-Declination diagram the galaxies from 
$12^{\rm h}15^{\rm m}< {\alpha}< 13^{\rm h}00^{\rm m}$, the strip being 
chosen to cut through Void 1 and Void 2 (Fig. 2\,a), containing thus
some of the void galaxies.  
The \lq\lq finger of God\rq\rq\ that is the
Coma Cluster is in fact at the border of the surveyed region, lying
mainly outside it. 

Fig. 3\,b contains the next strip between $13^{\rm h}00^{\rm m}< {\alpha}< 
13^{\rm h}45^{\rm m}$, plotted again in a redshift-Declination diagram.
The cone contains now the easter side of the Arch and cuts through the 
filaments that we mentioned as remnants of
the \lq\lq Harvard Sticky Man\rq\rq. 

In Fig. 3\,c we plotted one more redshift-Declination cone, for the strip in
Right-Ascension $13^{\rm h}45^{\rm m}< {\alpha}< 14^{\rm h}15^{\rm m}$.
The strip was chosen to cut through the filaments that runs in radial direction
in our Fig. 2\,a and therefore do not contain relevant nearby voids for our
surveyed region. 

In order to have a better impression of the whole surveyed region, we 
projected in a redshift-Right Ascension
diagram (Fig. 4) a strip of 15$^{\circ}$ in Declination, from 
$30^{\circ}30^{\prime}<{\delta}<45^{\circ}30^{\prime}$. Projection effects 
would of course affect the cone but in this case we are interested only to 
which extent the voids are still defined in the diagram. Due to the crowding 
of the diagram
we refrain from plotting the faint ZCAT galaxies, and we consider only the 
comparison catalogue (bright ZCAT galaxies). It is 
remarkable to see that despite the large strip in Declination that was
projected in the cone, the two isolated galaxies in Void 1 are
still clearly isolated and the void is still very well defined. This indicates
that the void  extends at least 15 degrees in Declination.
  
Overall the wedge diagrams show that our ELGs follow the structures traced by
the normal galaxies. However 17 ELGs (17$\%$) are very isolated, of which at 
least 8 ($8\%$) lie in some foreground voids. There are also some ELGs that lie
at the rim of the voids and there seem to be a tendency for the ELGs to 
be more evenly distributed that the ZCAT galaxies.

\subsection{The nearest neighbour test}

In order to quantify the visual impression provided by Figures 2-4, we applied
some statistical tests for differences in the distribution of HS and ZCAT
samples. For the ELGs we consider only the galaxies from the complete sample
derived in section $\S$ 2, namely the galaxies that have m$_{L+C}\le12.0$
(F$_{L+C}\ge 6.7\times 10^{-14} {\rm erg}\,{\rm sec}^{-1}\,{\rm cm}^{-2}$).

It is worth stressing that within the range of our surveyed regions
we deal with a field sample. No rich Abell cluster is present, only some
Zwicky clusters. In Region 3, the Coma Cluster
neighbours the southern border of our region (see Fig. 3\,a), with its main body 
outside. This implies that any differences that we could find will
not be due to the fact that the emission-line galaxies have the tendency to
avoid rich clusters.

We first applied a Kolmogorov-Smirnov test to the redshift distributions of the
two samples out to a velocity of 
10000\,km/s.  Since the velocity distribution of
the ZCAT falls off rapidly beyond 10000\,km/s, we limited our 
statistic to a subsample with velocities below this value. The comparison 
distribution was
constructed by selecting at random from the ZCAT the same number of 
objects (N) and in the same volume of space as in the ELG sample. This randomly
generated distribution was computed for 1000 samples of N ZCAT galaxies, and the
results averaged to produce the final comparison. The results indicate that the
two samples are drawn from the same parent population (KS=0.55).

\begin{figure}[htb]
\plotfiddle{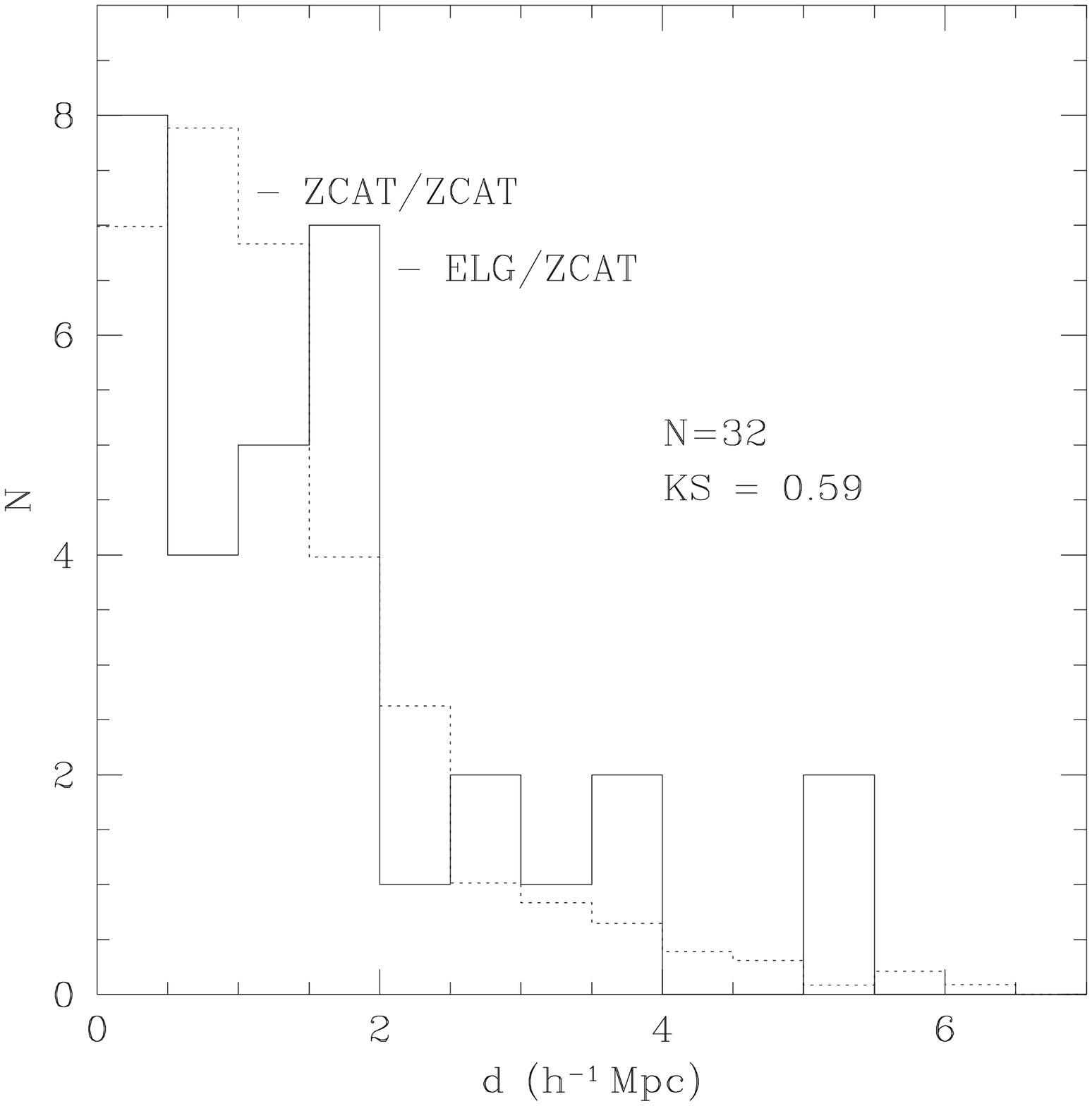}{4.0in}{0.}{40.}{40.}{-110}{-30}
{\baselineskip 1.5cm \caption[]{ The nearest neighbour distributions. The
ELG/ZCAT separations are plotted with solid line and the comparison ZCAT/ZCAT
distribution with dashed lines.   }}
\end{figure}

To better address the question of whether or not the two samples have the same
spatial distributions, we used a nearest neighbour (NN) test (Thompson 1983).
The cone-diagrams give also
an impression of the overall spatial structures, but as the plots are only two
dimensional representations, the projection effects could affect some of the
results. The nearest neighbour test calculates the real separation in the
3-dimensional space and also quantifies the results. Eder et al. (1989)
 showed that this test is particularly sensitive to the lack of clustering in 
a sample, and is therefore recommended for field samples. We 
limited again our statistic to subsamples with velocities less than 
10000\,km/s. We have computed two
distributions. One gives the separation between each ELG galaxy of the sample
(N objects) and the nearest ZCAT galaxy in the same field, but taken into
consideration the edge effects. This means that the ZCAT galaxies were taken 
from a
slightly larger field than that of the ELGs. For the second distribution we
followed the same procedure as for the redshift distribution: a randomly selected
sample of N galaxies was taken from the ZCAT catalogue. We calculated then the
separation between each of the N ZCAT galaxies and its nearest ZCAT neighbour,
again with edge effects considered.

The NN distributions are shown in Fig. 5. The overall impression is that the 
two distributions are quite similar. There seems to be an excess of ELGs at
intermediate separations, around 2\,h$^{-1}$Mpc, but the errors in each bin 
are quite big, due to the pure number statistics. There are also some ELGs at
 higher separations, where the ZCAT do not contribute. But these 
differences
cannot change the overall similarity between the two distributions. This is
confirmed by a Kolmogorov-Smirnov test which gives a KS=0.59, which means that
the two distributions are identical. We should notice that some of the very
isolated galaxies are not contained in the
complete sample and therefore were not included in the computation of the NN
test. These galaxies produce a tail of high separations in the
distribution of ELGs, which  
sharpen the difference between the two distributions and make the ELGs to be
more uniformly distributed than the giant galaxies.

The results of our statistical tests should be considered with caution, 
since the
comparison ZCAT is not complete up to 15.5. The incompleteness of the
comparison catalogue could introduce some errors that cannot be
controlled. Unfortunatelly, until the pulic release of the CfA2, the ZCAT is 
the only catalogue that samples the distribution of the normal galaxies on a 
large enough extent. 

\subsection{Discussion}

\begin{table*}[htp]
\caption[]{The main characteristics of the void galaxies. }
\begin{tabular}{ccccccccr}
                      & & & & & & & &\\
\hline\hline
                 & & & & & & & &\\
    name        & v & D$_{NN}$ & D$_{NN}$ & 
D$_{NN}$ & B &  M & Flux & EW\,\,\,\,\, \\
                &   & {\scriptsize (HS-ZCAT)} & 
{\scriptsize (HS-HS-ZCAT)} & {\scriptsize (HS-ZCAT(late))} & & & {\scriptsize 
[OIII] ${\lambda}$5007} & {\scriptsize [OIII] ${\lambda}$5007} \\
 & {\scriptsize km/s} & {\scriptsize h$^{-1}$\,Mpc} & 
{\scriptsize h$^{-1}$\,Mpc} & {\scriptsize h$^{-1}$\,Mpc} & & &
 {\scriptsize
erg\,sec$^{-1}\,{\rm cm}^{-2}$} & {\scriptsize ${\rm \AA}$}\,\,\,\,\,\,\,\, \\  
(1) & (2) & (3) & (4) & (5) & (6) & (7) & (8) & (9)\,\,\,\,\,\, \\
                 & & & & & & & & \\
\hline
                 & & & & & & & &\\
HS1342+3354 & 1692.5 & \,\,\,1.93 & 1.77 & \,\,\,2.50 & 19.2 & -12.73 
& 1.85e-14 & -271.49\\
HS1349+3942 & 1695.5 & \,\,\,2.20 & 1.77 & \,\,\,2.20 & 16.7 & -15.20 
& 3.07e-14 & -46.69 \\
HS1236+3821 & 2214.8 & \,\,\,3.85 & 3.85 & \,\,\,3.85 & 15.3 & -17.25 
& 3.38e-14 & -23.80 \\ 
HS1226+3719 & 3306.0 & \,\,\,6.63 & -    &  -    & 19.2 & -14.24 & 0        
& 0 \\
HS1236+3937 & 5571.4 & \,\,\,8.68 & 7.88 & \,\,\,8.68 & 18.7 & -15.88 
& 2.54e-14 & -466.52\\
HS1232+3947 & 6355.6 & \,\,\,5.26 & 3.97 & \,\,\,5.26 & 17.2 & -17.67 
& 3.80e-14 & -130.20\\
HS1240+3721 & 6582.3 & \,\,\,3.34 & 3.34 & \,\,\,3.34 & 17.9 & -17.05 
& 0 &     0 \\
HS1332+3426 & 6664.5 & \,\,\,4.45 & 2.06 & \,\,\,4.45 & 18.6 & -16.37 
& 1.97e-14 & -162.54 \\
HS1328+3424 & 6849.1 & \,\,\,2.76 & 2.06 & \,\,\,2.76 & 16.8 & -18.23 
& 9.36e-15 & -17.90\\ 
HS1310+3801 & 6954.1 & \,\,\,3.43 & 3.43 & \,\,\,3.43 & 18.1 & -16.97 
& - & - \\
HS1529+4512 & 7088.0 & \,\,\,6.83 & 4.45 & 10.20 & 17.9 & -17.22 
& 1.49e-14 & -94.35\\
HS1325+3255 & 7942.5 & \,\,\,3.46 & 3.46 & \,\,\,3.46 & 19.2 & -16.16 
& 2.39e-14 & -388.33\\
HS1306+3320 & 8131.0 & \,\,\,3.45 & 3.45 & \,\,\,3.58 & 16.9 & -18.51 
& 4.04e-14 & -108.82\\
HS1526+4045 & 8777.3 & \,\,\,3.59 & 3.59 & \,\,\,3.59 & 17.0 & -18.60 
& 5.40e-15 & -10.91\\
HS1341+3117 & 8846.4 & \,\,\,5.69 & 5.69 & \,\,\,5.69 & 18.6 & -17.00 
& 1.68e-14 & -80.30\\
HS1429+4511 & 9742.1 & 11.39    & 8.43 & 11.39 & 18.0 & -17.81 
& 2.45e-15 & -7.89\\
HS1507+3743 & 9776.2 & \,\,\,5.27 & 5.27 & \,\,\,8.21 & 17.9 & -17.93 
& 2.19e-13 & -1465.93\\
                & & & & & & & &\\
\hline
\end{tabular}
\end{table*}

In this section we discuss the significance of our findings in some of
the nearby voids. We refer only to the two nearest and
best defined voids presented in Figure 2a, Void 1 and Void 2. First we 
estimate how many
normal galaxies brighter than 15.5 we expect to find in the voids if the 
galaxies would be uniformly distributed. We consider the ZCAT galaxies 
brighter than 15.5 because the voids were defined by the distribution of these 
galaxies. We calculate the volume of Void 1
considering for simplicity an elipsoid shape with the diameters of the 
main axis: 3500\,km/s, 1$^{\rm h}30^{\rm m}$ and 15$^{\circ}$ (in radial
velocity, Righ Ascension and Declination). We took also into consideration 
that not all the volume of the void was surveyed. At a distance of z=0.01 this
 will give a volume of 1289\,h$^{-3}{\rm Mpc}^{3}$. If we integrate the 
luminosity function derived by de Lapparent et al. 
(1989), over the magnitude range that includes galaxies brighter than 15.5,
 one would expect to find 106 galaxies brighter
than 15.5. This result is obtained on the assumption that the 
galaxies were independently and randomly distributed, which is obviously not 
the
case. One could of course correct for the fraction of galaxies that are not
independent by using the autocorrelation function. For simplicity we consider
only our rough estimates and we obtained an underdensity of 106. For 
Void 2 we obtain a volume of 1793\,h$^{-3}{\rm Mpc}^{3}$ and an underdensity 
 of 57. This is the underdensity in the distribution of the normal 
galaxies. 

The number of ELG one would expect in the voids was calculated using 
the space densities derived in subsection $\S$ 3.1, $\Phi=0.011\,{\rm
Mpc}^{-3}$. We then estimate to find 44 ELGs in Void 1 and 47 ELGs in Void 2. 
This would be the case if our sample would be 100$\%$ complete over the 
magnitude range for which the space densities were considered. We already
mentioned that the incompleteness factor was 2.04, therefore we should expect
only  22 galaxies in Void 1 and 23 galaxies in Void 2. We found one ELG in Void
1 ( a second one is not an ELG, see $\S$ 4) and 7 ELGs in Void 2 (of which 2
are at the rim of the void), which means we did not find a significant void 
population at the density that was tested (the density of the walls and
filaments). Then the void population has either a density that is at least a
factor 4.5 lower, or alternatively, that the void population is even fainter 
than 
the limits of our survey, and we reached only the brightest peaks of such
population. One cannot reject the hypothesis that the few galaxies found in
voids do not form a population, rather they represent fluctuations of the large
scale structure. This would explain why some voids were found to contain a few
galaxies and other voids to be empty (Rosenberg et al. 1994, 
Pustil'nik et al. 1995).

\section{The properties of the void galaxies}

A natural question that would arise from a study that has the goal to search
for galaxies in voids is whether the galaxies found in the low density regions
have special properties in comparison with the characteristics of the total 
sample.
 We did not find a real void population, but we could still comment on 
the properties of the isolated galaxies found in some voids. Table 3 gives 
the main parameters of the void galaxies from the surveyed region   
analysed in this paper. The galaxies are ordered with respect of increasing 
the radial velocity, because the isolation of the galaxies also increases with 
distance. Column (1) contains the name of the galaxies and Column (2) gives 
the radial velocity. In the following columns we listed the separation between
 our galaxies and their nearest bright ZCAT neighbour, D$_{NN}$(HS-ZCAT) (Column
 (3)); between our galaxies and their nearest neighbour, D$_{NN}$(HS-HS-ZCAT) 
(which can be also one galaxy from our sample or a bright 
ZCAT) (Column (4)) and between our galaxies and their nearest late-type bright 
ZCAT neighbour, D$_{NN}$(HS-ZCAT(late)) (Column (5)). For one galaxy, 
HS1226+3719, which is a galaxy with absorption (see discussion below), only 
the separation to its nearest bright ZCAT is given. Then we give the apparent B
magnitudes (Column (6)), the absolute M$_{B}$ magnitude (Column (7)), the flux
 of the [OIII] ${\lambda}$5007 line (Column (8)) and the EW of the same line 
(Column (9)). For the galaxies for which the [OIII] ${\lambda}$5007 line is 
not 
detected, the fluxes and EW are set to 0. For one galaxy, HS1310+3801, the 
fluxes and EW are not available. This galaxy was already known in the 
literature and it was
not observed by us, and therefore was also not included in the statistical 
analyse. 

The separations of the void galaxies show that 50$\%$ have the
nearest neighbour among themself. This would suggest that the void galaxies
are not uniformly distributed but also have the tendency to form fainter
structures inside the bigger voids. Unfortunately, the low number of our 
isolated galaxies cannot allow us to draw any definitive
conclusion. Nevertheless, some fainter ZCAT galaxies have been
found to associate with some of our void galaxies. For example the Arch is
followed by four faint ZCAT galaxies, and from these, all are late types. In
Table 4 we list the ZCAT galaxies we found in Voids 1 and 2, respectively (even
though they do not come from the complete sample),
together with their main parameters (velocity, B magnitude and T morphological
type, when available). The Arch seems also to divide Void 2 in three smaller
voids. Lindner et al. (1996) and Szomoru et al. (1996b) also suggested that the
galaxies found in voids have the tendency to cluster. In addition, 75$\%$ of 
our void galaxies have
the nearest ZCAT neighbour among the late type galaxies. This should not come
as a surprise since 
the late type galaxies have the tendency to be less clustered than the early 
types, which would preferentially form the skeleton of the clusters. 

\begin{figure*}[htb]
\plotfiddle{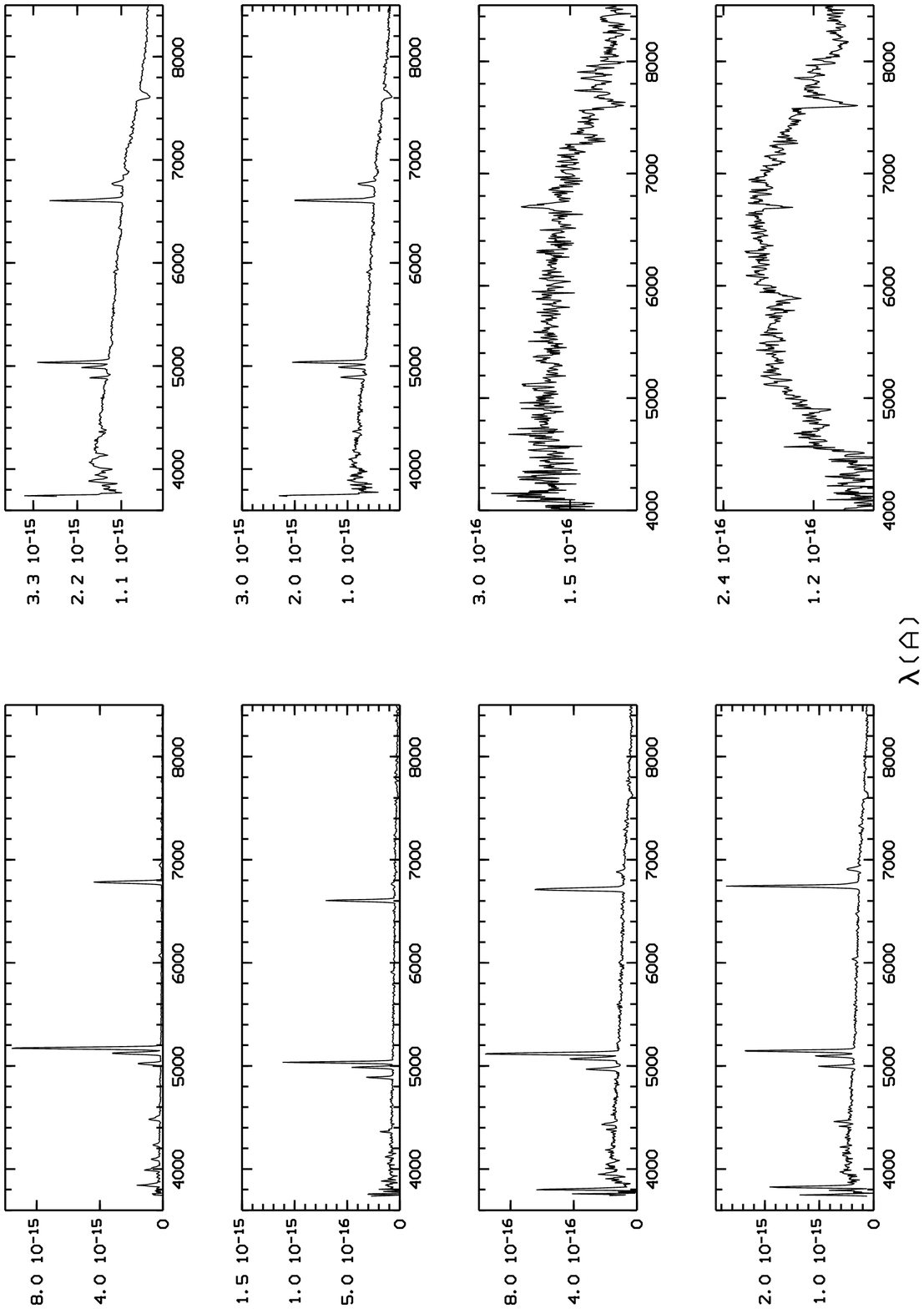}{4.0in}{-90.}{50.}{50.}{-185}{280}
\caption[]{Spectra of void galaxies. The y-axes contain the fluxes per 
unit of wavelength, F$_{\lambda}$ 
(erg\,sec$^{-1}\,{\rm cm}^{-2}\,{\rm \AA}^{-1}$) while the x-axes contain the
wavelength, ${\lambda}$ (\AA). }
\end{figure*}

\begin{table}[htp]
\caption[]{Void galaxies in the ZCAT sample fainter than B=15.0.}
\begin{tabular}{cccccc}
& & & & &\\
\hline\hline
& & & & &\\
 & ZCAT & other  & v & B & T\\
 & name         & name           & (km/s) & & \\
& & & & &\\
\hline
& & & & &\\
Void 1 & 1149+3510 & MK641 & 2165 & 16.5 & \\
       & 1240+3541 &   -   & 1806 & 17.0 & \\
& & & & &\\
\hline
& & & & &\\
Void 2 & 1227+3311 & CG1027 & 5606 & 15.7 & \\
& 1230+3952 & PGC41579 & 6212 & 16.0 & Sc\\
       & 1305+3307 & PGC45451 & 5300 & 16.0 & Sd\\
       & 1305+3302 & PGC45481 & 6753 & 16.0 & Sd\\
& & & & &\\
\hline
\end{tabular}
\end{table}

The absolute magnitudes of our void galaxies range between 
$-12.7<{\rm M}_B<-18.5$
  which means that all of them are dwarfs.
These galaxies are therefore intrinsically faint objects, quite
different from the galaxies that were found in the Bootes void (Weistrop et
al. 1995), which were
mainly M$^{*}$ galaxies or brighter. Nevertheless, the distribution of 
absolute magnitudes
of our parent sample starts to drop around M$_B=-17$ and only a few galaxies 
are found in the range $-12<M_B<-15$. If a faint void population had 
typical luminosities around M$_B=-15$ and below, we would just start to detect
 it, since we are very incomplete at the faint end. The few galaxies we found 
in the voids could constitute the tip of the iceberg of the void population. 

The spectroscopic properties of our void galaxies are very different and 
they do not belong to only one class of objects. The EW and fluxes encompass 
mostly the whole range of values from the parent sample. Some void
galaxies have extremely large EW and fluxes of the [OIII] ${\lambda}$5007 line 
whether others are barely detectable. In Figure 6 we give some example of
void galaxy spectra. The plots show 
that the galaxies we found in voids have different degrees of ionization, from 
very high ionization objects with very faint continuum, close to the extreme
case of Searle-Sargent objects, up to very low ionization galaxies and strong
continuum which indicate an underlying older stellar population. Amazingly, we 
found also one object that has no detectable [OIII] ${\lambda}$5007 line;
the only emission-line being H${\alpha}$. This 
object was
selected mainly because of the blue continuum and was considered  a
second priority candidate (see Paper 1 for a detailed description of the 
selection procedure) and was not included in the statistical analyse. But the 
most unexpected void galaxy is HS1226+3719, an object that entered in our 
sample as a
failure of our selection procedure, being a galaxy with absorption. 
On the 
other hand this galaxy is one of our best void candidates, with an isolation  
of ${\rm D}_{NN}=6.63h^{-1}$Mpc, and lying in the centre of Void 1.
(Figure 2\,a). This result fact makes 
us wonder whether a population of dwarf elliptical galaxies would not be in
fact the 
hidden void population which would recover the biasing theories. But it is 
known (Binggeli 1989) that the dwarf ellipticals are the most clustered galaxies in
 the Universe, which populate mainly the clusters. This still do not exclude 
the possibility that the voids would be occupied by a population of dwarf red 
galaxies similar to the dwarf ellipticals we see now in clusters. 

\section{Summary and conclusions}

Our present study did not find the voids occupied by a homogenous population 
of dwarf galaxies. We found a few galaxies in the very well
defined nearby voids, but the number of void galaxies is not significant at the
density level of field galaxies. We could interpret this result in the 
sense that a void population, if any, should have the density at least a factor
4.5 lower than the density of walls and filaments. Another possibility is 
 that we start to see the brightest peaks of such population and we are still 
not faint enough to really sample these objects. Alternatively, there is no 
void population,
 and the galaxies we found 
in voids were only fluctuations from the normal distribution. This would
explain why
some voids were found to be empty and other to contain a few galaxies. On the
other hand we should also remember that we were probing the voids only with 
emission-line galaxies, and from these, we are mainly sensitive in the high 
ionization objects. We are complete in the galaxies with large 
equivalent widths and we start to miss objects that have fainter EW. These 
objects are mainly low ionization and are better detected by H${\alpha}$
 surveys, like the surveys based on objective prism IIIaF plates (UCM survey, 
Zamorano et al. (1994)). 
Therefore the voids were still not probed by low ionization emission-line
objects. Nevertheless the UCM galaxies seem also to follow the same structures
as the normal galaxies (Gallego 1995), but the UCM sample do not contain 
galaxies as faint as we detect. We should also keep in mind that the
IIIaJ selected emission-line galaxies contribute only 7$\%$ from the total
number of galaxies (Salzer 1989). It was suggested that the small HII galaxies
are the best candidates to fill the voids, but it could still be that a
population of old red dwarf galaxies would fill the voids. These galaxies do
not have a strong spectroscopic signature like an emission spectrum and they are
 gas poor  and therefore are not detected by HI surveys. If they were
fainter than M$_B$=-15, they would be not properly sampled by any survey that 
search them on direct images.

Coming back to our ELGs we suggest that the few galaxies that we found
in voids have also a tendency for clustering. This result is far from
being secure, but some of our void galaxies seem to associate with 
faint late-type ZCAT galaxies. For example our Arch of 7 ELGs in
Void 2 is followed by four faint ZCAT galaxies. The Arch 
seem to divide that bigger void in 3 smaller voids. And not at the end, 
75$\%$ from
our isolated galaxies have their nearest neighbours among themself. These
results are close to some results of N-body simulations (Dubinski et al. 1993,
van de Weygaert \& van Kampten 1993) or with the observational results of an
HI survey in the Bootes void (Szomoru 1996). It was thus suggested that smaller
scale voids disappear within larger voids but the frozen-in remnants of small
walls would produce smaller substructures inside the larger voids. Lindner et 
al. (1996) also suggest a hierarchical  distribution of galaxies and voids, in the 
sense that superclusters and clusters delimit bigger voids, bright galaxies 
some smaller voids and faint galaxies delimit very small voids, inside the 
bigger ones.

Overall the main characteristics of the ELG spatial distributions are:

1. The ELGs are better than normal galaxies for tracing the luminous matter 
at further distance;

2. The ELGs have a small tendency to be more evenly distributed than the ZCAT
galaxies, with some galaxies lying in some voids or at the rim of the voids.

3. A filamentary structure (the Arch), populated only by faint ELGs, has been 
found to cross a big void in front of the Great Wall.

4. The void galaxies are intrinsically faint and they do not have special
spectroscopic properties in comparison with the other ELGs in the field.

\begin{acknowledgements}

We would like to thank Dr. A.P. Fairall for the careful review of this
manuscript and to Dr. H. Hagen for his contribution during the work with
the HS data base. We also thank Dr. S. Pustil'nik for providing us some 
unpublished data on Markarian galaxies. We gratefully acknowledge the comments
 and discussions with Drs. B. Binggeli, A. Burkert, R. P Kirshner, 
D. Thompson, D. Valls-Gebaud. We would also like to thank the Calar Alto staff
for their support during the observations. U. Hopp acknowledge the support by 
the SFB 375 by the Deutsche Forschungsgemeinschaft.

\end{acknowledgements}

\end{document}

%% file: eps.tex
\def\eps@scaling{.95}
\def\epsscale#1{\gdef\eps@scaling{#1}}

\def\plotone#1{\centering \leavevmode
    \epsfxsize=\eps@scaling\columnwidth \epsfbox{#1}}

\def\plotfiddle#1#2#3#4#5#6#7{\centering \leavevmode
    \vbox to#2{\rule{0pt}{#2}}
    \includegraphics{#1}}